\documentstyle[preprint,prl,aps]{revtex}
\begin{document}
\title{Competing Patterns of Signaling Activity \\ in
{\it Dictyostelium discoideum}}
\author{Kyoung J. Lee$^1$, Edward C. Cox$^2$, and Raymond E. Goldstein$^1$}
\address{$^1$Department of Physics, Joseph Henry Laboratories,
Princeton University, Princeton, NJ 08544}
\address{$^2$Department of Molecular Biology, Princeton University,
Princeton, NJ 08544}

\maketitle

\begin{abstract}

Quantitative experiments are described on spatio-temporal
patterns of coherent chemical signaling activity in
populations of {\it Dictyostelium discoideum} amoebae.
We observe competition between
spontaneously firing centers and rotating spiral waves that depends
strongly on the overall cell density.  At low densities, no complete spirals
appear and chemotactic aggregation is driven by periodic concentric waves,
whereas at high densities the firing centers seen at early times
nucleate and are apparently entrained by spiral waves whose
cores ultimately serve as aggregation centers.
Possible mechanisms for these observations are discussed.

\end{abstract}

\pacs{PACS numbers:   87.10.+e, 87.22.-q, 82.40.Ck, 47.54.+r}

In a variety of contexts in the biological world we find coherent
spatio-temporal patterns of propagating chemical waves \cite{Rensing}.
Often, as in cardiac tissue, these waves are triggered and globally
controlled by specialized cells termed ``pacemakers" \cite{cardiac}.
In other systems, traveling chemical waves may arise by a process of
self-organization of undifferentiated cells.
A well-known biological example is provided by populations of the slime mold
{\it Dictyostelium discoideum} \cite{dicty}, which upon nutrient deprivation
sustain waves of cyclic adenosine 3$^{\prime}$,5$^{\prime}$-monophosphate
(cAMP)
that drive chemotactic migration of cells.
These waves result from
a relay mechanism within each cell, and travel by means of the diffusive
coupling of nearby cells through the substrate.
Despite a large body of experimental and
theoretical work on the coupled dynamics of wave propagation and associated
chemotaxis leading to cell aggregation \cite{dicty,Levine}, the
means by which a particular pattern of coherent traveling waves emerges from
the non-signaling state has remained unclear.  Many studies have
revealed the presence of concentric waves (or ``target" patterns)
emanating from periodically firing pacemakers \cite{dicty,pacemakers},
while others have focused on the rotating spiral waves that also
occur \cite{Seigert,Tyson}, and are qualitatively similar
to those seen in chemical systems such as the Belousov-Zhabotinski (BZ)
reaction \cite{BZ,Meron}.  The distinction between these two types of
patterns is significant; targets require an autonomous pacemaker,
while rotating spirals do not.  Yet, there has been no clear experimental
determination of the factors controlling which of these two signaling
patterns dominates.

We report here a quantitative experimental study of the competition
between autonomous firing centers and rotating spiral waves
in {\it D. discoideum}.
Using cell population density as a control parameter, we find that
when the cell density is high spiral waves dominate at late times,
whereas at low cell density the asymptotic pattern is dominated
by circular waves emanating from pacemakers, as shown in Fig. 1.
In addition, the spiral waves themselves originate in the instabilities of
finite wave segments \cite{Durston_theory} and
apparently entrain the firing centers.
In a low density system, the firing centers also entrain each other.
In both cases, the dynamical evolution proceeds from random firings to
{\it periodic} events.
The density-dependent competition between signaling patterns is
similar to that seen recently in the BZ reaction in the presence of
a catalyst imprinted on a lattice of varying density \cite{Showalter}.
The appearance of spirals only in a limited range of densities suggests
that variations in the diffusive coupling of excitable elements lead to
pattern selection. Our results also relate to recent theoretical
studies of mechanisms for spiral symmetry-breaking
\cite{Aranson}, specific biochemical origins for pattern
evolution \cite{Palsson}, and feedback mechanisms \cite{Levine_nature},
as well as the role of stochasticity in pattern selection \cite{Jung}.

First, we briefly review the essential features of the signaling
mechanism. Signaling begins with the synthesis of the messenger molecule
cAMP from ATP by adenylate
cyclase within individual cells \cite{kin_ref}.  It is also
degraded to 5$^{\prime}$-AMP by phosphodiesterase. Both cAMP and
phosphodiesterase are excreted through the cell membrane wall.  The
production of cAMP within individual cells is stimulated and
controlled by the state of cAMP receptors in the cell membrane:
when the receptors are fully saturated with cAMP, the synthesis of
internal cAMP stops.  It is now well documented both in experiments
and model studies \cite{Tyson} that these chemical reactions can
produce rotating spiral waves and circular
waves of collective chemical activity in spatially-extended systems
like those shown in Fig. 1.

In our experiments, cells were grown using standard culturing
techniques \cite{culture}.  This preparation yields a monolayer
of cells spread on the surface of an agar layer in a dish.  Signaling
is generally initiated within several hours after nutrient deprivation,
during which time the cells were kept in the dark.
Patterns of signaling activity were monitored using a dark field
optical setup \cite{dark_field_details}, sensitive to changes in optical
properties of the cells in response to chemical waves \cite{Devreotes}.

Figure 1 shows the evolution of spatio-temporal
patterns observed in experiments with two different initial cell
densities.  In both cases, the spontaneous firing of many cells is observed
in the form of spreading circular wavefronts.
Spiral waves can form from broken wave segments that naturally arise from
inhomogeneities in the medium.  When the initial cell density $\rho$ is high
($\rho_H$),
as in Fig. \ref{image_evol}(a), competition between
circular waves and spirals is quite apparent  -- spirals
eventually suppress all firing events.

A dramatically different situation arises when the initial cell
density is low ($\rho_L$), shown in Fig. \ref{image_evol}(b).  As in the
system with $\rho_{H}$, many firing centers appear at an
early stage (but with a longer induction period), and, at the same
time, wave segments also form from inhomogeneities in the medium.
In contrast to high density populations, however, the broken ends of the waves
do not evolve to full spirals, and firings continue to the later cell
aggregation phase.

We have located
all the firing events occurring during the signaling phase of development.
Focusing first on the case of high cell density,
Figure \ref{composite}(a) shows the positions
of firing centers along with those
of the spiral cores that ultimately form.  We find that
most centers fire only once.  Given our spatial resolution, the size of
the cells ($\sim 8$ $\mu$m), and
the scale of their random motion, it is not possible for us to
determine whether those most closely-spaced clusters of data
points arise from distinct cells firing independently, or instead from
identical cells  which have simply moved during the course of observation.
Since spirals evolve from the ends of broken waves, the
locations of the spiral cores are not correlated with the centers.

A quantification of firing-center/spiral competition is the time evolution of
the number of firing events.  Figure \ref{N}(a) shows this for the
same $\rho_H$ as Fig. \ref{composite}(a).
The number $N$ of firing events initially increases in time, reaches a
maximum value, then ultimate vanishes as the spirals become more fully formed.
Let us now contrast this behavior with our observations at low densities.
First, as shown in Fig. \ref{composite}(b), there are overwhelmingly many more
firing centers at $\rho_L$ than $\rho_H$
[compare with Fig. \ref{composite}(a)].
With the smaller average distance between centers, the ability of
broken wave segments to survive as spirals is greatly diminished.

At $\rho_L$, no complete spirals form, and firing centers survive
until the later aggregation stage (at $\approx 11$ hours),
as shown in Fig. \ref{N}(b).
Figure \ref{N} illustrates the fact that the
fractional subpopulation
of firing centers ($N/\rho$) {\it decreases} with density.
This suggests that the formation of
a center is not simply a characteristic of a subpopulation of cells, but
rather is determined in part by cell-cell interactions \cite{Glazer}.
Likewise, the lack of correlation between the positions of the centers and
the spiral cores suggests that the latter are determined primarily by the
positions and dynamics of wave segments, rather than intrinsic properties
of the cells at the core.

Systems with several different $\rho$ in addition to the two already discussed
have also been studied.  In a system with $\rho
= 1.5\times\rho_{L}$, similar entrainment dynamics appear between
spirals and centers and between different centers.  In this case, $N$
initially increases, reaches a maximum, decreases gradually as before,
but to a nonzero value, as shown in Fig. \ref{N}(c).  In systems with
higher values of $\rho$ (toward and beyond $\rho_{H}$), spirals
gradually extinguish firing events with the same
entrainment dynamics as $\rho_{H}$.

At all densities, a peak in the number of firing centers
occurs at intermediate times.
[The gradual increase in $N$ during the period
10-12 hours at $\rho_L$ is due to the increase in the
firing {\it frequency} of the surviving centers.]
In the high-density systems, the decrease in $N$ beyond the peak
appears to be due to a  suppression of firing centers by spiral waves.
Some direct evidence for this
is presented in Fig. \ref{blownup}.  Over the course of $3$ periods of
firing activity, a pair of nearby spiral waves repeatedly passes through
a center (indicated by an arrow). Because the firing frequency of the
center is slower than
the rotation frequency of spiral waves \cite{Gross}, the
collision zone at which waves
annihilate gradually moves toward the center (first three frames in
Fig. \ref{blownup}), and ultimately the firing center is suppressed (last frame
in Fig. \ref{blownup}).
This observed behavior can be viewed as an {\it entrainment} of slow centers
by faster periodic spiral waves.
A corollary to this
observation is that since the spiral waves all have very similar
rotation frequencies they are not effective in suppressing one another.
However, we have observed some examples of the breaking of symmetry between
members of a spiral pair, leading to dominance of one over the other
(as seen left-of-center in the first frame of Fig. 1).
This is consistent with theoretical arguments based
on the existence of certain ultra-slow chemical dynamics \cite{Aranson},
or from chemical inhomogeneities in the medium \cite{Palsson}.
Similar observations on the suppression of slow autonomous
centers by spirals were reported without quantification
in earlier experiments on
the Belousov-Zhabotinskii reaction in a Petri dish \cite{BZ_exp} and
in cellular automaton models \cite{BZ_model}.
The decrease in $N$ in the low-density systems appears to arise from
the competition {\it among} firing centers, rather than from the influence of
incomplete spirals (wave segments).  For instance, a center that fires
with a higher frequency may suppress a slowly firing neighbor.
Taken together, these
earlier observations and the present ones suggest that this
entrainment phenomenon is a generic feature of excitable media.

Among the possible mechanisms
for firing center suppression, several appear quite likely on the
basis of known experimental and/or theoretical observations.
First, from the general phenomenology of excitable media, we may
expect the passage of cAMP waves past a center to act as a kind of
phase resetting event, leading to
synchronization.  This would
be a {\it local} version of oscillator entrainment,
mediated by the passage of waves, complementary to that studied
recently in systems with {\it global} coupling \cite{fire,joseph}.
It may also be an example of ``spatio-temporal stochastic
resonance" \cite{Jung}, in which a wave of excitation
entrains noisy excitable elements (e.g. the firing centers).
Second, since the thickness of the agar
substrate may play a role in the diffusive coupling of
neighboring cells \cite{dicty}, gradual changes in the local
chemical environment of a cell by the repeated passage of waves may alter
its excitability.
Third, gradual changes in the internal properties of the cells
(e.g. density of membrane receptors, etc.) during the signaling stage
may also play a role \cite{Aranson,Palsson}.

Future experiments to elucidate the mechanism of
selection of signaling activity will focus on controlled
perturbations of the system.  One avenue of interest involves
external stimulation through changes in background levels of chemicals
such as cAMP in the agar substrate (both static and periodic in time).
A second, complementary to our use of cell density, involves the use of
different wild-type and mutant strains with variations in biochemical
properties associated with signaling \cite{Kessin}.
Experiments along these lines are currently in progress.
Finally, the interplay between chemotaxis and pattern selection is
an important area for further experimental and theoretical study, perhaps
along the lines of recent work on bacterial systems \cite{Budrene}.

We acknowledge many helpful discussions with I. Aranson, R.H. Austin,
J.T. Bonner, and H. Levine.  The work was supported in
part by an NSF Presidential Faculty Fellowship (DMR 93-50227) and the
Alfred P. Sloan Foundation (REG),
and NSF IBN 93-04849 (ECC).

\begin{figure}
\caption{Dark field images showing the evolution of two kinds
of signaling activity.  In (a) spirals dominate
circular waves at high density [$\rho_H=21.8\times10^{5}$
(cells/cm$^{2}$)], whereas in (b) circular waves dominate
[$\rho_L=7.3\times10^{5}$ (cells/cm$^{2}$)].  Time is
indicated in hours and minutes elapsed from the point of food
deprivation. Each 24 mm $\times$ 18 mm image was obtained by subtracting
two successive
images taken 30 sec. apart and rescaling the resulting image
over 256 grey levels to enhance the contrast.  The bright bands and
the accompanying dark shadows indicate the cells in an active state,
while the grey background corresponds to inactive cells. \label{image_evol}}
\end{figure}

\begin{figure}
\caption{Composite diagrams showing the location of firing
centers (filled circles) and spiral cores (empty circles): (a)
$\rho=\rho_{H}$; (b) $\rho=\rho_{L}$.
 \label{composite}}
\end{figure}

\begin{figure}
\caption{Dynamical evolution of number of firing
centers per unit time (min) and unit area (cm$^{2}$): (a)
$\rho=\rho_{H}$; (b) $\rho=\rho_{L}$; (c) composite of (a) and (b)
and $1.5\times\rho_{L}$ (triangles),
and $3.5\times\rho_{L}$ (diamonds).  The plots were obtained by counting
the number of firing centers in $15$ minute intervals.
\label{N}}
\end{figure}

\begin{figure}
\caption{Enlarged images showing suppression of a
low-frequency center by a higher frequency spiral wave at $\rho_H$.
\label{blownup}}
\end{figure}

\narrowtext

\end{document}